
----------
X-Sun-Data-Type: default
X-Sun-Data-Name: Prl97vrwd.tex
X-Sun-Charset: us-ascii
X-Sun-Content-Lines: 416

\documentstyle[preprint,aps]{revtex}

\begin{document}
\title{Is the magnetic field necessary for the Aharonov-Bohm effect in mesoscopics ?}
\author{I.D. Vagner$^{1}$, A.S. Rozhavsky$^{1,2}$, P. Wyder$^{1}$ and A.Yu. Zyuzin$%
^{1,3}$}
\address{$^{1}$Grenoble High Magnetic Field Laboratory\\
Max-Planck-Institut f\"{u}r Festk\"{o}rperforschung and\\
Centre National de la Recherche Scientifique,\\
BP 166, 38042 Grenoble CEDEX 09, France.\\
$^{2}$ B.I. Verkin Institute for Low Temperature Physics and Engineering,\\
47 Lenin Avenue, 310164, Kharkov, Ukraine.\\
$^{3}$ A.F. Ioffe Physico-Technical Institute,\\
Russian Academy of Sciences, 194021, St. - Peterburg, Russia.}
\maketitle

\baselineskip=2.7ex

\smallskip

\newpage \abstract{We predict a new class of topological mesoscopic
phenomena in absence of external magnetic field, based on combined action of
the nonequilibrium nuclear spin population and charge carriers spin-orbit
interaction (meso-nucleo-spinics). We show that Aharonov-Bohm like
oscillations of the persistent current in GaAs/AlGaAs based mesoscopic rings
may exist, in the absence of the external magnetic field, provided that a
topologically nontrivial strongly nonequilibrium nuclear spin population is
created. This phenomenon is due to the breaking, via the spin-orbit
coupling, of the clock wise - anti clock wise symmetry of the charge
carriers momentum, which results in the oscillatory in time persistent
current. }


\newpage

Persistent currents (PC) in mesoscopic rings reflect the broken clock
wise-anticlock wise symmetry of charge carriers momenta caused, usually, by
the external vector potential. Experimentally PCs are observed when an
adiabatically slow time dependent external magnetic field is applied along
the ring axis \cite{Levy90,Chandr91,Mailly91}. The magnetic field variation
results in the oscillatory, with the magnetic flux quantum $\Phi _{0}=\frac{%
hc}{e}$ (or its harmonics) period behavior of the diamagnetic moment (the
PC), which is the manifestation of the Aharonov-Bohm effect (ABE).

We propose here that in a quantum ring with a nonequilibrium nuclear spin
population the persistent current will exist, even in the absence of
external magnetic field. We predict the ABE like oscillations of PC with
time, during the time interval of the order of nuclear spin relaxation time $%
T_{1}$ \cite{SlichterBk}, which is known to be long in semiconductors at low
temperatures \cite{DP84}. The hyperfine field , caused by the nonequilibrium
nuclear spin population \cite{DP84} breaks the spin symmetry of charged
carriers. Combined with a strong spin - orbital (SO) coupling, in systems
without center of inversion \cite{PZ95,BR84} , it results in the breaking of
the rotational symmetry of diamagnetic currents in a ring. Under the
topologically nontrivial spatial nuclear spin distribution, the hyperfine
field produces an adiabatically slow time variation of the Berry phase of
the electron wave function, analogous to one which emerges in standard ABE
in textured mesoscopic rings \cite{LGB90,BKKR91} and in Aharonov-Casher
effect \cite{BA93}. The time variation of this topological phase results in
observable oscillations of a diamagnetic moment (the persistent current). We
emphasize that this is one of a series of ''meso-nucleo-spinic'' effects,
which may take place in mesoscopic systems with broken, due to the combined
action of the hyperfine field and spin-orbital interaction, symmetry.

The hyperfine interactions in GaAs heterojunctions and similar quantum Hall
systems attracted recently sharply growing theoretical \cite{VM88} and
experimental \cite{Berg90,Wald94,Barrett95} attention. The main recent
physical interest in this subject is based on a fact that the discrete
nature of the electron spectrum in these systems will result in
exponentially long: T$_{1}\sim $ $\exp \{\Delta /T\}$ (here $\Delta $ is the
electron energy level spacing and $T$ is the temperature) dependence of the
nuclear spin relaxation times $T_{1}$ on the system parameters, \cite{VM88}.
We assume here that similar law should take place also in the nanostructures
with well defined size quantization of the electron spectrum. Note that in
this case $T_{1}$ is very sensitive to the potential fluctuations, caused by
the inhomogeneous distribution of impurities in a heterojunction \cite{IMV91}
. Intensive experimental studies \cite{Berg90,Wald94,Barrett95} of this
phenomenon have provided a more detailed knowledge on the hyperfine
interaction between the nuclear and electron spins in heterojunctions and
quantum wells. It was observed that the nuclear spin relaxation time is
rather long (up to 10$^{3}\sec $) and the hyperfine field acting on the
charge carriers spins is extremely high, up to 10$^{4}G$ \cite{Berg90,Wald94}%
. Till now, however, the manifestations of the hyperfine interaction in
quantum interference (mesoscopic) phenomena has not been studied.

Spin-orbit interaction plays an important role in mesoscopic physics. In
disordered quantum rings it was studied in \cite{MGE-W89,E-WGMO92}, where
universal reduction of the PC harmonics was obtained. Similar conclusions
were drawn in \cite{CM95} where the SO interaction specific for a
GaAs/AlGaAs heterojunctions was studied. We show here that in such systems
SO combined with the hyperfine field, produced by strongly nonequilibrium
nuclear spin population can play a constructive role in creating an analog
to AB effect.

The contact hyperfine interaction is\cite{SlichterBk}:

\begin{equation}
\widehat{H}_{chf}^{en}=\frac{8\pi }{3}\mu _{B}\gamma _{n}\hbar ^{2}\sum_{i}%
{\bf I}_{i}\cdot {\bf \sigma }\delta ({\bf r}-{\bf R}_{i}).  \label{Hchf1}
\end{equation}
Here $\mu _{B}$ is the Bohr magneton, $\gamma _{n}$ is the nuclear magneton, 
${\bf I},{\bf \sigma },{\bf R}_{i},{\bf r}$ are the nuclear and the charge
carriers spins and position vectors, respectively. It follows from Eq. (\ref
{Hchf1}), that once the nuclear spins are polarized, i.e. if $\left\langle
\sum_{i}{\bf I}_{i}\right\rangle \neq 0$ , the charge carriers spins feel
the effective, hyperfine field $B_{hypf}=B_{hypf}^{o}\exp \left(
-t/T_{1}\right) $ which lifts the spin degeneracy even in the absence of
external magnetic field. In GaAs/AlGaAs one may achieve the spin splitting
due to hyperfine field of the order of the one tenth of the Fermi energy 
\cite{Berg90,Wald94}.

Let us suppose therefore that the charge carriers spin orientation is
partially polarized during the time interval of the order of $T_{1}$ . It is
quite obvious that the topologically nontrivial spin texture combined with
the spin-orbit interaction will result in a persistent current.

The spin-orbit interaction in GaAs/AlGaAs heterostructures was widely
studied both theoretically and experimentally (see \cite{PZ95} and
references therein). The main contributions came from a) bulk inversion
assymmetry and b) from the structure inversion asymmetry, first pointed out
in \cite{BR84} . Since both contributions are of comparable value, in what
follows we will concentrate, for simplicity, on the typical for a
heterojunction Bychkov-Rashba term \cite{BR84}
\begin{equation}
\widehat{H}_{so}=\frac{\alpha }{\hbar }\sum_{i}\left[ {\bf \sigma }%
_{i}\times {\bf p}\right] {\bf \nu ,}  \label{HsoRashba1}
\end{equation}
where $\alpha =0.6\cdot 10^{-9}eVcm$ for holes with $m^{*}=0.5m_{0}$ ($m_{0}$
is the free electron mass)\cite{BR84,SSCTGW83}, and $\alpha =0.25\cdot
10^{-9}eVcm$ for electrons \cite{BR84,SKW83}, ${\bf \sigma }_{i}$ ,${\bf p}%
_{i}$ are the charge carrier spin and momentum and ${\bf \nu }$ is the
normal to the surface. It can be rewritten in the form

\begin{equation}
\widehat{H}_{so}={\bf pA}_{eff},  \label{Hso=pAeff}
\end{equation}
where 
\begin{equation}
A_{eff}^{GaAs}\simeq \frac{\alpha m^{*}}{\hbar }\left\langle \sigma
\right\rangle ,  \label{AeffGaAs}
\end{equation}
$\left\langle \sigma \right\rangle $ stands for a nonequilibrium carriers
spin population. Under the conditions of a topologically nontrivial
orientation of ${\bf A}_{eff}^{GaAs}$ (see the discussion below) the wave
function of a charge carrier encircling the ring gains the phase shift
similar to the one in an external magnetic field like in the ordinary ABE.
This phase shift can be estimated as follows

\begin{equation}
2\pi \Theta =\frac{1}{\hbar }\oint A_{eff}^{GaAs}dl=\frac{m^{*}}{\hbar ^{2}}%
\left\langle \sigma (t)\alpha \right\rangle \sim \frac{m^{*}\sigma (t)\alpha 
}{\hbar ^{2}}L,  \label{phi1}
\end{equation}
where $L$ is the ring perimeter. To observe the oscillatory persistent
current connected with the adiabatically slow time-dependent $\left\langle
\sigma \left( t\right) \right\rangle ,$ $L$ is supposed to be less than the
phase breaking length. Taking the realistic values for $L\approx 3\mu m$ 
\cite{Mailly91} and $\left\langle \sigma \right\rangle \approx 0.05\div 0.1$%
\cite{Berg90,Wald94} we estimate $2\pi \Theta \sim 5\div 10$ which shows the
experimental feasibility of this effect.

The standard definition of the spontaneous diamagnetic current is

\begin{equation}
j_{hfso}=-c\frac{\partial F}{\partial \phi }\mid _{\phi _{ext}=0},
\label{jhfso}
\end{equation}
where F is the electron free energy and $\phi _{ext}$ is the external
(probe) magnetic flux. The oscillations of persistent current arise due to
the exponential time dependence of the phase $\Theta _{eff}^{0}\exp \left\{
-t/T_{1}\right\} $ in Eq. (\ref{phi1}), with the time constant $T_{1}.$From
the analogy with the standard ABE in a mesoscopic ring, we expect the
following form for a persistent current in an one dimensional quantum ring
at low enough (T $\ll \Delta $) temperature

\begin{equation}
j_{hfso}\sim \frac{ev_{F}}{L}\sin (2\pi \Theta _{o}e^{-\frac{t}{T_{1}}}).
\label{jhfso1}
\end{equation}
Here $\Theta _{o}$ is the initial phase value. We outline the marking
difference between the periodical time dependence of standard AB
oscillations, which are observed usually under the condition of linear time
variation of the applied magnetic field and the hyperfine driven
oscillations which die off due to the exponential time dependence of the
nuclear polarization.

In what follows, we present the microscopic justification of the qualitative
model discussed above.

The microscopic description is based on a following Hamiltonian

\[
\widehat{H}=-\frac{\hbar ^{2}}{2m^{*}\rho ^{2}}\left( \frac{\partial }{%
\partial \varphi }-i\frac{\Phi }{\Phi _{o}}\right) ^{2}-i\frac{\alpha
(\varphi )}{\rho }\left( \sigma _{z}\cos \varphi +\sigma _{y}\sin \varphi
\right) \left( \frac{\partial }{\partial \varphi }-i\frac{\Phi }{\Phi _{o}}%
\right) 
\]

\begin{equation}
+i\frac{\alpha (\varphi )}{2\rho }\left( \sigma _{z}\sin \varphi -\sigma
_{y}\cos \varphi \right) -\frac{i}{2\rho }\frac{d\alpha }{d\varphi }\left(
\sigma _{z}\cos \varphi +\sigma _{y}\sin \varphi \right) -\mu
_{B}B_{hypf}\sigma _{z}.  \label{Hqm}
\end{equation}
It describes the carriers, confined to an one-dimensional ring of radius $%
\rho $ , placed in the (ZY) plane, in a case of inhomogeneous SO coupling
which, as will be discussed later, is one of the possible realizations of a
topologically nontrivial effective vector potential Eq. (\ref{AeffGaAs}).
The ring is pierced by an external magnetic flux $\Phi $ , $B_{hypf}$ is the
hyperfine field oriented along the $Z$ -axis. Note, that the Hamiltonian,
Eq. (\ref{Hqm}), is a generalization of $\alpha =const$, studied in \cite
{CM95} , for the case $\alpha =\alpha \left( \varphi \right) $.

Let us show, within the perturbation theory, that the l.h.s. of the Eq. (\ref
{jhfso}) is nonzero at $\Phi =0$ . Since the electron Zeeman splitting in
GaAs/AlGaAs, caused by the hyperfine field, $B_{hypf}$ at strongly
nonequilibrium nuclear population, exceeds $\alpha \hbar /\rho $ in micron
rings, we may safely neglect the effective spin-orbital field compared to
the hyperfine one. The zero-approximation eigenfunctions and eigenvalues
are, respectively

\[
\Psi _{n}^{\left( \pm \right) }=\frac{1}{\sqrt{2\pi }}\left( 
{1 \overwithdelims() 0}%
,%
{0 \overwithdelims() 1}%
\right) \exp \left( in\varphi \right) ; 
\]
\begin{equation}
\varepsilon _{n}^{\left( \pm \right) }=\frac{\hbar ^{2}}{2m^{*}\rho ^{2}}%
\left( n-\frac{\Phi }{\Phi _{o}}\right) ^{2}\pm \mu _{B}B_{hypf}.
\label{PSI,eps}
\end{equation}

The flux-dependent first order corrections to $\varepsilon _{n}^{\pm }$ are

\begin{equation}
V_{nn}^{\pm }=\pm \frac{1}{4\pi \rho }\left( n-\frac{\Phi }{\Phi _{o}}%
\right) \int_{0}^{2\pi }d\varphi \alpha \left( \varphi \right) \cos \varphi
=\pm \left( n-\frac{\Phi }{\Phi _{o}}\right) \frac{\left\langle \alpha
\right\rangle }{\rho }.  \label{Vnn+-}
\end{equation}

It follows that each level carries current if the integral in the r.h.s. of
the Eq. (\ref{Vnn+-}) is nonzero. This implies topological restrictions on
the function $\alpha \left( \varphi \right) $which are equivalent to
existence of a topologically nontrivial spin texture.

Calculation of the first order correction to the free energy is
straightforward, and using the Eq. (\ref{jhfso}) the spontaneous persistent
current can be found to be

\begin{equation}
j_{hfso}\simeq \frac{c\Delta }{\rho \Phi _{0}T}\left\langle \alpha
\right\rangle \sum_{n=0}^{\infty }n^{2}\left\{ \cosh ^{-2}\left( \frac{%
n^{2}-\mu _{-}}{T^{*}}\right) -\cosh ^{-2}\left( \frac{n^{2}-\mu _{+}}{T^{*}}%
\right) \right\} ,  \label{jhfso3}
\end{equation}

here $\Delta =\hbar ^{2}/2m^{*}\rho ^{2}$ ;$\mu _{\pm }=\left( \mu \pm \mu
_{B}B_{hypf}\right) /\Delta $ , where $\mu $ is the chemical potential, $%
T^{*}=T/\Delta $ .

At low temperatures: T$\ll \Delta \ll \mu _{B}B_{hypf}\ll \mu $ , the Eq. (%
\ref{jhfso3}) reads

\begin{equation}
j_{hfso}\simeq \frac{2ev_{F}\left\langle \alpha \right\rangle \left\langle
\sigma \right\rangle m^{*}}{\hbar ^{2}}\sim \frac{ev_{F}}{L}\Theta \left(
t\right)  \label{jhfso5}
\end{equation}

where

\begin{equation}
\left\langle \sigma \right\rangle =\frac{\mu _{B}B_{hypf}}{\mu }
\label{<sigma>3}
\end{equation}
is the nonequilibrium spin population introduced in the Eq. (\ref{AeffGaAs}%
), and $\Theta \left( t\right) $ is the topological phase Eq. (\ref{phi1}).
We note, that Eq. (\ref{jhfso5}) is the first term of the expansion of the
phenomenological equation Eq. (\ref{jhfso1}) . It follows, that the direct
quantum mechanical calculation confirms the existence of a spontaneous
persistent current, induced by combined action of the nonequilibrium nuclear
spin population and spin-orbit coupling.

Consider now the possibility of creation of a topologically nontrivial
effective vector potential defined in Eq. (\ref{phi1}) . In the geometry
where the vector ${\bf \nu }$ , Eq. (\ref{HsoRashba1}), is normal to the
heterojunction, either the nuclei polarization driven $\sigma \left( \varphi
\right) $ or the spin-orbit coupling $\alpha \left( \varphi \right) $should
be inhomogeneous along the perimeter of the ring. Out of a variety of
different experimental realizations let us outline the following ones.

Since the nuclear relaxation rate T$_{1}^{-1}$ in GaAs/AlGaAs heterojunction
is highly sensitive to the impurities distribution \cite{IMV91}, one can
achieve different nuclear polarization in different parts of a mesoscopic
ring, provided the characteristic length of the impurity potential is of
order of several hundreds of $A$ , i.e. comparable to the ring width. In
this configuration we expect that the mesoscopic sensitivity to a single
impurity position may produce a nonvanishing phase Eq. (\ref{phi1}).

Another way of obtaining a nonzero circulation of the effective vector
potential is creation of the slowly varying on the scale $k_{F}^{-1}$
coordinate dependent spin-orbit coupling $\alpha \left( \varphi \right) $%
connected with external potentials like boundaries, heavy atoms impurities
along the perimeter of the ring and other imperfections which may locally
modify the spin-orbit interaction in these systems.

To summarize, we propose here a new, hyperfine field driven
(meso-nucleo-spinic) mesoscopic effect : the Aharonov-Bohm like oscillations
of a persistent current in a GaAs/AlGaAs mesoscopic ring in the absence of
external magnetic field. We note that the large (of the order of 1Tesla)
hyperfine field results in the nonequilibrium population of the charge
carriers spins. The latter, under the conditions of the topologically
nontrivial spatial effective vector potential distribution results in a
persistent current. This current is oscillating (aperiodically) and
decreasing with time during the time interval of the order of the nuclear
spin-relaxation time $T_{1}$. At low temperatures $T_{1}^{-1}$ can be
sufficiently long and the predicted oscillations experimentally observable..

We acknowledge illuminative discussions with H. Bednarski, Yu. Bychkov, A.\
Dyugaev, V. Gurevich, I. Krive, G. Kventzel', T. Maniv, L. Levy and B.
Spivak.

I.V. acknowledge the support by US-Israeli Binational Science Foundation
grant No. 94-00243.

\end{document}